\documentclass[12pt]{article}
\usepackage{graphicx}
\usepackage{amsmath}
\usepackage{hhline}
\usepackage{amssymb}
\usepackage{times}
\usepackage{cite}
\usepackage{array}
\usepackage{multirow}
\usepackage{color}
\definecolor{rltred}{rgb}{0.75,0,0}
\definecolor{rltgreen}{rgb}{0,0.5,0}
\definecolor{rltblue}{rgb}{0,0,0.75}
\newif\ifpdf
\ifx\pdfoutput\undefined
    \pdffalse          
\else
    \pdfoutput=1       
    \pdftrue
\fi

\ifpdf
\usepackage{thumbpdf}
\usepackage[pdftex,
        colorlinks=true,
        urlcolor=rltblue,       
        filecolor=rltgreen,     
        linkcolor=rltred,       
        pdftitle={Example File for pdflatex},
        pdfauthor={H1},
        pdfsubject={Template file},
        pdfkeywords={High-Energy Physics, Particle Physics},
        pdfpagemode=None,
        bookmarksopen=true]{hyperref}
\fi

\newlength{\dinwidth}
\newlength{\dinmargin}
\newcommand{\jpsi}{$J/\psi$}

\newcommand{\wgp}{W_{\gamma p}}

\begin{document}  
\newcommand{\qsq}{\ensuremath{Q^2} }
\newcommand{\gevsq}{\ensuremath{\mathrm{GeV}^2} }
\newcommand{\et}{\ensuremath{E_t^*} }
\newcommand{\rap}{\ensuremath{\eta^*} }
\newcommand{\gp}{\ensuremath{\gamma^*}p }
\newcommand{\dsiget}{\ensuremath{{\rm d}\sigma_{ep}/{\rm d}E_t^*} }
\newcommand{\dsigrap}{\ensuremath{{\rm d}\sigma_{ep}/{\rm d}\eta^*} }
\newcommand{\ra}{\rightarrow}
\newcommand{\psip}{$\psi(2S)$}
\newcommand{\ccbar}{\ensuremath{\mit c\overline{c}}}

\def\Journal#1#2#3#4{{#1} {\bf #2} (#3) #4}
\def\NCA{\em Nuovo Cimento}
\def\NIM{\em Nucl. Instr. Methods}
\def\NIMA{{\em Nucl. Instr. Methods} {\bf A}}
\def\NPB{{\em Nucl. Phys.}   {\bf B}}
\def\PLB{{\em Phys. Lett.}   {\bf B}}
\def\PRL{\em Phys. Rev. Lett.}
\def\PRD{{\em Phys. Rev.}    {\bf D}}
\def\ZPC{{\em Z. Phys.}      {\bf C}}
\def\EJC{{\em Eur. Phys. J.} {\bf C}}
\def\CPC{\em Comp. Phys. Commun.}

\begin{titlepage}

\begin{flushleft}
DESY 03-061 \hfill ISSN 0418-9833\\
June 2003
\end{flushleft}  

\noindent

\vspace{2cm}

\begin{center}
\begin{Large}

{\bf \boldmath Diffractive Photoproduction of ${J/\psi}$  Mesons \\
with Large Momentum Transfer \\ at HERA}

\vspace{2cm}

H1 Collaboration

\end{Large}
\end{center}

\vspace{2cm}

\begin{abstract}
\noindent
The diffractive photoproduction of $J/\psi$ mesons
is measured with the H1 detector  at the $ep$ collider HERA 
using an integrated luminosity of 78 pb$^{-1}$.
The differential cross section ${\rm d}\sigma(\gamma p \rightarrow J/\psi Y)
/{\rm d}t$
is studied in the range  $2 < |t| < 30 \rm ~GeV^{2}$,
where $t$ is the square of the four-momentum transferred at the
proton vertex. 
The cross section is also presented
as a function of the photon-proton centre-of-mass
energy $\wgp$ in three $t$ intervals,  spanning the range
$50 < \wgp < 200 \ {\rm GeV}$. 
A fast rise 
of the cross section with  $\wgp$ is observed 
for each $t$ range 
and the slope for the effective linear
Pomeron trajectory is measured to be
$\alpha^{\prime}= -0.0135 \pm 0.0074 \ {\rm (stat.)} 
\pm 0.0051 \ {\rm (syst.)}~{\rm GeV^{-2}}$.
The measurements are compared with
perturbative QCD models based on BFKL and DGLAP evolution. 
The data are found to be compatible with $s$-channel helicity
conservation.
\end{abstract}

\vspace{1.5cm}

\begin{center}
Submitted to {\em Phys. Lett. B.} 
\end{center}

\end{titlepage}

\begin{flushleft}


A.~Aktas$^{10}$,               
V.~Andreev$^{24}$,             
T.~Anthonis$^{4}$,             
A.~Astvatsatourov$^{35}$,      
A.~Babaev$^{23}$,              
S.~Backovic$^{35}$,            
J.~B\"ahr$^{35}$,              
P.~Baranov$^{24}$,             
E.~Barrelet$^{28}$,            
W.~Bartel$^{10}$
% ...

\bigskip{\it
 $ ^{1}$ I. Physikalisches Institut der RWTH, Aachen, Germany$^{ a}$ \\
% ...
$ ^{4}$ Inter-University Institute for High Energies ULB-VUB, Brussels;
          Universiteit Antwerpen (UIA), Antwerpen; Belgium$^{ c}$ \\
% ...
$ ^{10}$ DESY, Hamburg, Germany \\
% ...
$ ^{23}$ Institute for Theoretical and Experimental Physics,
          Moscow, Russia$^{ l}$ \\
 $ ^{24}$ Lebedev Physical Institute, Moscow, Russia$^{ e}$ \\
% ...
$ ^{28}$ LPNHE, Universit\'{e}s Paris VI and VII, IN2P3-CNRS,
          Paris, France \\
% ...
$ ^{35}$ DESY, Zeuthen, Germany \\
% ...
$ ^a$ Supported by the Bundesministerium f\"ur Bildung und Forschung, FRG,
      under contract numbers 05 H1 1GUA /1, 05 H1 1PAA /1, 05 H1 1PAB /9,
      05 H1 1PEA /6, 05 H1 1VHA /7 and 05 H1 1VHB /5 \\
% ...
$ ^c$ Supported by FNRS-FWO-Vlaanderen, IISN-IIKW and IWT \\
% ...
$ ^e$ Supported by the Deutsche Forschungsgemeinschaft \\
% ...
$ ^l$ Partially Supported by Russian Foundation
      for Basic Research, grant    no. 00-15-96584 \\
\smallskip
% ...
}

\end{flushleft}

\newpage
\section{Introduction} 
\label{sect:intro}

The diffractive photoproduction of \jpsi\ mesons 
with large negative
momentum transfer squared $t$ 
at the proton vertex
is a powerful means to probe the parton 
dynamics of the diffractive exchange.
The variable $t$ provides a relevant 
scale to investigate the application of perturbative QCD (pQCD).
The diffractive photoproduction of vector mesons can be 
modelled in the proton rest frame, where the photon 
fluctuates into a $q\bar{q}$ pair 
at a long distance from the proton target. 
The colour singlet 
exchange between the $q\bar{q}$ fluctuation and the proton is 
realised in lowest order QCD by the
exchange of a pair of gluons with opposite colour. 
In the leading logarithmic (LL) approximation, this process
is described by the effective exchange of a gluonic ladder.
At sufficiently low values of 
Bjorken $x$ (i.e. large values of the centre-of-mass energy $\wgp$),
the gluon ladder 
is expected to include contributions from
BFKL evolution~\cite{FKL:1976},  
as well as from  
standard DGLAP evolution~\cite{Gribov:ri}.
Compared with other channels which have been 
used to search for BFKL 
evolution~\cite{Adloff:1998fa,Adloff:1999zx,Adloff:2002em,Derrick:1995pb,Enberg:2001ev,Cox:1999dw},
the measurement of diffractive \jpsi\ production at large $|t|$
provides an experimentally clean signature 
in which the accurate measurement of the \jpsi\ four-momentum allows the 
kinematic dependences of the process to be determined precisely.

In this paper, an analysis of the diffractive photoproduction process
$\gamma p \rightarrow J/\psi Y $ is presented, extending into the
hitherto unexplored region of large $|t|$ ($2 < |t| < 30$ \rm {GeV}$^2$).
Here, the system $Y$ represents either an elastically 
scattered proton or a dissociated proton system.  
For the range of $|t|$ studied in this analysis, 
the contribution from elastic $J/\psi$ production
may be neglected due to its steep $|t|$ dependence~\cite{Adloff:2002re}.
The cross section is measured differentially as a function of $|t|$ 
and as a function of the photon-proton centre-of-mass energy $\wgp$
in different regions of  $|t|$, using the $J/\psi$ decay 
into two muons.
To obtain information about the helicity structure of the 
interaction, 
the spin density matrix elements are extracted.

\section{Perturbative QCD Models} \label{sect:models}

Perturbative QCD models
for the photoproduction of \jpsi\ mesons
have been developed 
in the leading logarithmic approximation using either 
BFKL~\cite{Forshaw:1995ax,Bartels:1996fs,Forshaw:2001pf}
or DGLAP~\cite{Gotsman:2001ne} evolution.
In the BFKL LL model the cross section 
depends linearly on the parton 
distribution of the proton and the gluon ladder couples to a single parton 
(dominantly a gluon) within the proton.
The BFKL amplitude is expanded in terms of 
$\log ({x_h \wgp^2/W^2_0})$, where 
$x_h$ is the fraction of the proton momentum carried
by the parton struck by the diffractive exchange.
The scale parameter $W_0$ is chosen to be half the vector meson
mass $M_V$.
The value of ${\alpha}_s$ is fixed in the model to a value consistent
with that extracted from a fit~\cite{Forshaw:2001pf} to 
proton dissociative $\rho$, $\phi$  and 
\jpsi\ photoproduction data at HERA~\cite{Chekanov:2002rm}.
The BFKL LL model predicts an approximate power-law behaviour for the 
$t$ dependence of the form ${\rm d} \sigma /{\rm d}t \propto |t|^{-n}$, where
$n$ is a function of $|t|$.
For the kinematic range studied here, $n$ increases from around $3$ to $4$
with increasing $|t|$ and 
the approximation to a power-law improves as $|t|$ increases.
The calculation predicts a fast rise of the cross section 
$\sigma \sim \wgp^{\delta}$
with $\delta \sim 1.4$, which has little or no dependence on the value of $t$.
In a recent paper \cite{Enberg:2002zy}, the LL calculations  have been 
extended to incorporate the effects of higher conformal
spin~\cite{Motyka:2001zh}.
Although the full next-to-leading order terms of the BFKL amplitude
have yet to be calculated for non-zero $t$,
an estimate of the 
non-leading (NL) corrections
was obtained using kinematic constraints.
In the DGLAP LL model, the cross section depends on
the squared gluon distribution of the proton.
The model predicts a non-exponential $t$ dependence and a steep 
energy dependence which flattens as $|t|$ approaches $M_V^2$
due to the limited phase space
available for evolution.  

In the pQCD models 
\cite{Forshaw:1995ax,Bartels:1996fs,Forshaw:2001pf,Gotsman:2001ne,Enberg:2002zy}, 
a non-relativistic approximation~\cite{Ryskin:1992ui} for  the $J/\psi$ 
wavefunction is used in which 
the longitudinal momentum of the vector meson
is shared equally between the quark and the anti-quark.
In this approximation, the vector meson retains the helicity
of the photon such that
$s$-channel helicity conservation (SCHC) is satisfied~\cite{Kuraev:1998ht}.

\section{Data Analysis} \label{sect:procedure}

The data presented here were recorded in the years 1996 to 2000 and
correspond to an integrated luminosity of $78 \:\mbox{pb}^{-1}$.
The majority of the data were collected when HERA was operated with 
positrons of energy $27.5~{\rm GeV}$ and protons of $920~{\rm GeV}$.
These data are combined with smaller data samples in which either the 
proton energy was $820~{\rm GeV}$ or the positrons were replaced 
by electrons.

\subsection{The H1 Detector} \label{sect:detector}

A detailed description of the H1 detector can be found in \cite{Abt:1997xv} 
and only a short overview of the detector components most 
relevant to the present analysis is given here. 
The $z$-axis of the H1 detector is defined along the beam direction
such that positive $z$ values correspond to the direction of 
the outgoing proton beam.

Charged particles emerging from the interaction region
are measured by the central tracking detector (CTD) in the pseudorapidity
range $-1.74 < \eta < 1.74$\footnote{The pseudorapidity
$\eta$ of an object detected with polar angle $\theta$ is defined as
$\eta = - \ln \ \tan (\theta / 2)$.}. The CTD comprises two large
cylindrical central jet drift chambers (CJC) and two $z$-chambers 
arranged concentrically around the beam-line 
within a solenoidal magnetic field of 1.15 T. 
The CTD also provides triggering
information  based on track segments in the $r-\phi$ plane 
from the CJC and the $z$-position of the vertex from a double 
layer of multi-wire proportional chambers.
The energies of final state particles are measured in the liquid 
argon (LAr) calorimeter, which surrounds the tracking chambers and 
covers the range $-1.5 < \eta < 3.4$. 
The backward region ($-4.0 < \eta < -1.4$) is covered by
a lead--scintillating fibre calorimeter (SPACAL~\cite{Nicholls:1996di}) with 
electromagnetic and hadronic sections.  
The calorimeters are surrounded by the iron return yoke of the solenoidal 
magnet.
The tracks of muons which penetrate the main detector are reconstructed from  
streamer tubes placed within the iron in the range 
$-2.5 < \eta < 3.4$.
The luminosity is measured using the small angle Bremsstrahlung process
($ep\rightarrow ep\gamma$) in which the final state photon is
detected in a calorimeter, close to the beam-pipe, at
$103 \ {\rm m}$ from the nominal interaction point.

\subsection{Kinematics} \label{sect:kinematic}

The kinematics for diffractive charmonium production 
$ep \rightarrow e J/\psi Y$
are described 
in terms of the $ep$ centre-of-mass-energy squared 
$s=(k+p)^2$, the 
virtuality of the photon \linebreak
$Q^2=-q^2=-(k-k')^2$,
the square of the centre-of-mass energy of the initial photon-proton system 
$\wgp^2 = (q+p)^2$
and the four-momentum transfer squared 
$t=(p-p_Y)^2$. Here $k$ ($k'$) is the four-momentum
of the incident (scattered) lepton 
and $q$ is the four-momentum of the virtual photon.
The four-momentum of the incident proton is denoted by $p$ 
and $p_Y$ is the four-momentum of the system $Y$.
The event elasticity is defined as 
$z = (p \cdot p_{\psi})/(p \cdot q)$ where $p_{\psi}$ is
the four-momentum of the \jpsi\ .
In the proton rest frame $z$ is equal to the fractional 
energy of the photon transferred to the vector meson.

\subsection{Event Selection} \label{sect:selection}

In this analysis, the $J/\psi$ mesons are detected via their decay 
into two oppositely charged muons 
(branching fraction $5.88 \pm 0.10\%$ \cite{Hagiwara:fs}).
The data were selected by 
a combination of triggers
based on muon and track signatures.
The selected events are required to have a 
vertex located in $z$ within $40~\rm cm$ of the nominal interaction point.
Events with two tracks of opposite charge in the CJC, 
each associated with the event vertex and each with
pseudo-rapidity $|\eta| < 1.74$ and transverse
momentum $p_T > 0.8 \rm ~GeV$ are used to form
$J/\psi$ candidates.
Both decay muons are identified in the instrumented iron
or as minimum ionising particles in the LAr calorimeter.

Photoproduction events are selected by the absence of a scattered 
beam lepton candidate in the LAr or SPACAL calorimeters.  
The accepted
photoproduction event sample covers the range \linebreak
$Q^2 \lesssim 1\, {\rm GeV}^2$ with 
an average $\langle Q^2\rangle \sim 0.06 \ {\rm GeV}^2$, as determined from
Monte Carlo simulations.

In order to select diffractive events, the analysis is restricted to the region
of elasticity $z > 0.95$.  
For the range of $t$ and $\wgp$ 
studied in this paper, the cut $z > 0.95$ restricts the invariant
mass of the system $Y$ to be in the range $M_Y \lesssim 30 ~{\rm GeV}$,
through the relation $z \simeq 1 - (M_Y^2 - t)/\wgp^2$.
The measurement of $z$ is obtained from 
$(E-p_z)_{J/\psi}/\sum (E-p_z)$ where 
$\sum (E-p_z)$ is calculated 
from all detected particles in the calorimeters and the CJC including
the decay products of the
$J/\psi$.
The variable $\wgp$ is reconstructed using 
$\wgp^2 = \sum (E-p_z) \cdot 2E_p$ where $E_p$ is 
the energy of the incident proton beam.
In the kinematic region studied, the variable $t$ is well
approximated by the negative transverse momentum squared of
the vector meson, i.e.\,$t\simeq -p_{t,J/\psi}^2$.

\subsection{Monte Carlo Simulation} \label{sect:acceptance}

Monte Carlo simulations are used to correct the data for the 
effects of resolution, acceptance and efficiency losses. 
Samples of events from signal and background
processes are passed through a detailed simulation of the
detector response, based on the GEANT program \cite{Brun:1987ma}, and 
through the same reconstruction software as was used for the data.

The Monte Carlo generator used for the simulation of proton dissociative 
diffractive \jpsi\ production
is HITVM~\cite{hitvm}, 
which generates events according to the 
BFKL model described in \cite{Forshaw:1995ax,Bartels:1996fs}.
The events are generated using the GRV94-HO parton 
density functions~\cite{Gluck:1994uf}    
and the partonic system is fragmented according to the
Lund string model implemented within
the JETSET program~\cite{jetset}. 
The generated $M_Y$ distribution in HITVM has an approximate exponential
dependence ${\rm d} \sigma / {\rm d}M_Y \sim e^{-0.1M_Y} $. SCHC is assumed 
for the photon to vector meson transition. 

The final sample of events contains background from 
resonant and non-resonant sources.
The resonant background
is produced indirectly through the decay of $\psi(2S)$ mesons.
This contribution is simulated using a Monte Carlo sample 
of $\psi(2S)$ mesons 
generated using the
DIFFVM Monte Carlo generator~\cite{List:jz} according to the $\psi(2S)$
$t$ distribution and cross section ratio to $J/\psi$ 
production measured at lower values of $|t|$~\cite{Adloff:2002re}.
A contribution of $4\%$ is observed with no significant $t$ dependence.
The main contribution to the non-resonant 
background is from the QED $\gamma\gamma \rightarrow \mu\mu$ process,
which is simulated using the LPAIR \cite{Baranov:1991yq} Monte Carlo 
generator.

\subsection{Signal Extraction}

The invariant mass spectrum  for all events 
in the range
$|t| > 2 \ {\rm GeV^2}$, $50 < \wgp < 150 ~{\rm GeV}$
and $z > 0.95$ is shown in figure~\ref{fig:signal}.
The LPAIR non-resonant background is normalised to the data in the side-bands 
outside the 
mass regions of the $J/\psi$ and $\psi (2S)$ resonances.
The number of signal events is determined from the
number of events in the
mass window of $ 2.9 < M_{\mu^{+} \mu^{-}} < 3.3~{\rm GeV}$,
after subtracting the contributions of the resonant 
and non-resonant backgrounds.
The resulting number of $J/\psi$ candidate events
for the total sample shown in figure~\ref{fig:signal}
is $846 \pm 30 \ {\rm (stat.)}$.

\subsection{Comparison of Data and Simulation}

The HITVM model gives a reasonable description of the data which is further 
improved through small adjustments to the $\wgp$ and $t$ distributions.
After these adjustments a comparison between 
the simulation and the 
data, before background subtraction, is given in figure~\ref{fig:control}
for the region 
$ |t| > 2 ~{\rm GeV^2}$, 
$50 < \wgp < 150 ~{\rm GeV}$, $z > 0.95$
and $2.9 < M_{\mu^{+} \mu^{-}} < 3.3 \ {\rm GeV}$.
Distributions are
shown for the polar angle and transverse momentum of
the decay muon tracks, 
for the reconstructed value of the elasticity
$z$ (where the cut on $z$ is not applied), for  
$\wgp$,  for the decay 
angular distributions
$\cos \theta^*$ and $\phi^*$ (see section~\ref{sect:spin})
and for the squared transverse momentum of the dimuon system
$p_{t,\mu^{+} \mu^{-}}^2$.
The structure in the $\phi^*$ distribution (figure~\ref{fig:control}f)
is due to the low acceptance for one of the muons, which has
a low transverse momentum in the laboratory frame, when the \jpsi\ meson 
production and decay planes coincide 
($\phi^* \sim 0^{\rm o}$ or $\phi^* \sim \pm 180^{\rm o}$). 

\subsection{Systematic Uncertainties}

The uncertainties in detector effects and in the modelling of the
underlying physics processes contribute to the systematic 
uncertainties in the cross section measurements. 
The following sources of systematic error are taken into account.

\begin{itemize}

\item
The uncertainty in the acceptance corrections is estimated 
by reweighting the $\wgp$ distribution
by $\wgp^{\pm 0.35}$ and the $t$ distribution 
by $t^{\pm 0.85}$.
The resulting systematic uncertainties on the cross section
measurements range from $1\%$ to $5\%$.

\item 
The uncertainty in the mass distribution of the proton dissociative
system $Y$ is estimated by reweighting the $M_Y$ dependence in HITVM
by $e^{\pm 0.06M_Y}$.
This results in a variation of the cross section 
of about $4\%$, increasing up to $19\%$ at the largest $\wgp$ and $|t|$.

\item 
The effect of possible deviations from SCHC is estimated
by modifying the simulated $\cos \theta^*$ distribution. 
The cross sections alter by $5\%$ on average.

\item
The uncertainty on the trigger efficiency, obtained from 
an independently triggered sample of events, gives a contribution 
to the systematic error of $6\%$. 

\item
The uncertainty in the identification efficiency of muons
is estimated by detailed comparison of the
data and simulation efficiencies for an
independent data sample.
The resulting systematic uncertainty is $6 \%$. 

\item
The uncertainty due to the 
reconstruction efficiency of the 
central tracker for the two tracks
leads to an error of $4\%$. 

\item
The uncertainty in the non-resonant background subtraction
is estimated by using a data side-band subtraction as an alternative
to the Monte Carlo subtraction. 
A difference of $\sim2\%$ is found
between the two methods and assigned to the systematic error.

\item 
The uncertainty in the subtraction of the 
$\psi (2S)$ background leads to an error of $2\%$, obtained by 
varying the normalisation and exponential $t$ slope of the $\psi (2S)$ cross
section in the simulation.

\item
Other sources of systematic error are the uncertainty 
in the hadronic energy scale of the liquid argon calorimeter, 
the uncertainty in the luminosity measurement 
and the uncertainty in 
the branching fraction for the measured decay channel  
\cite{Hagiwara:fs}.
Each of them is responsible for an error of no more than $1.7\%$.

\end{itemize}

The total systematic error for each data point has been obtained by adding
all individual contributions in quadrature. It has a small
dependence on $t$
with an average value of $12\%$
and increases from around $11 \%$ at low  $\wgp$ to $20\%$ at high 
$\wgp$.  
The part of the uncertainty which is uncorrelated between different 
data points 
contributes $8.5\%$ to the systematic error.
The statistical error 
is larger than the systematic error
in  the region  $|t| \ge 5.5 \ {\rm GeV^2}$.

\section{Results}
\label{sect:results}

\subsection{Cross Sections}

The differential cross section ${\rm d}\sigma  /{\rm d}t$ for the
process ${e p \rightarrow e J/\psi Y}$ is obtained 
from the number of data events in each measurement interval 
after corrections for backgrounds and detector effects, 
divided by the integrated luminosity, the branching fraction and the 
width of the interval.
The cross section for the
photoproduction process ${\gamma p \rightarrow J/\psi Y}$ is 
obtained by dividing the differential $ep$ cross section 
by the effective photon flux~\cite{Budnev:de} 
integrated over the $\wgp$ and $Q^2$ ranges
of the measurement.
QED radiative effects are estimated to be less than $1\%$ and
are neglected.
The differential photoproduction cross section ${\rm d}\sigma  /{\rm d}t$ 
is shown in figure~\ref{fig:t} and table~\ref{tab:dsigdt}
for the kinematic region $ 50 < \wgp < 150 ~{\rm GeV}$ 
and $z>0.95$.
The data are plotted at the mean value in each $t$ interval 
according to a parameterisation of the data.
In the region $|t| > 3.45 \ {\rm GeV^2}$,  the data in figure~\ref{fig:t}
are adequately described by a
power-law dependence of the form
$A \cdot {|t|^{-n}}$ where  
$n = 3.00 \pm 0.08 \ {\rm (stat.)} \pm 0.05 \ {\rm (syst.)}$.
When the power-law fit is repeated, each time increasing
the starting value of $|t|$ in the fit, 
the value of $n$ is found to increase systematically
up to a value of 
$n = 3.78 \pm 0.17 \ {\rm (stat.)} \pm 0.06 \ {\rm (syst.)}$
for $|t| > 10.4 \ {\rm GeV^2}$.
The data are incompatible with an exponential behaviour
${\rm d}\sigma / {\rm d}t \propto e^{bt}$  which was found to give a 
reasonable description of the proton dissociative \jpsi\  cross
section at lower values of $|t|$ ($ |t| < 5 \ {\rm GeV^2})$~\cite{Adloff:2002re}.

In figure~\ref{fig:t} the data are compared with the predictions from 
pQCD calculations in the BFKL leading 
logarithmic approximation~\cite{Enberg:2002zy} (solid curve), 
including non-leading corrections with fixed $\alpha_s$~\cite{Enberg:2002zy}
(dashed curve)
and including non-leading corrections with 
running $\alpha_s$~\cite{Enberg:2002zy } (dotted curve).
The $t$ dependence and normalisation 
of the data are well described by the
BFKL LL approximation 
when the parameters of the model are
set to values consistent with those extracted from a 
fit~\cite{Forshaw:2001pf} to various vector meson 
proton dissociation data at HERA
covering a smaller $|t|$ range~\cite{Chekanov:2002rm},
i.e. the scale parameter is set to $W_0=M_V/2$ 
and ${\alpha}_s$ is fixed at $0.18$.
The normalisation uncertainty due to the choice of $W_0$ is large. 
For example, using $W_0=M_V/4$ ($W_0=M_V$)  
leads to an increase (decrease) in the normalisation of the
prediction by a factor of approximately two.
The inclusion of NL corrections with a fixed strong coupling $\alpha_s$  
leads to only a small difference with respect to the LL
prediction.  However, with a running $\alpha_s$ the $t$ dependence
becomes steeper and the prediction is unable to describe the data
across the whole $t$ range.
The uncertainties in the
choice of the scale parameter, proton parton density and other parameters 
used in the NL calculation have only a small effect on the shape
of the predictions in comparison to the treatment of $\alpha_s$.  
The data are also compared with calculations in the DGLAP LL 
approximation~\cite{Gotsman:2001ne} (dashed-dotted curve) 
in the region of validity for the model $|t| < M_{J/\psi}^2$.
The data are well described in shape and normalisation 
when the separation parameter $t_0$,
which represents the value of $t$ at which the prediction for proton 
dissociation matches the elastic cross section, is set 
to $-0.60 \ {\rm GeV^2}$. 

The ZEUS collaboration has recently published data on the
diffractive production of $J/\psi$ mesons with proton 
dissociation in the range 
$1.2 < |t| < 6.5 ~{\rm GeV^2}$, $80 < \wgp < 120 ~{\rm GeV}$
and $x_h = |t|/(\wgp^2(1-z)) > 0.01$ \cite{Chekanov:2002rm}.
When the present analysis is performed in this kinematic region, good 
agreement between the H1 and ZEUS results is observed. 

In figure~\ref{fig:w} and tables~\ref{tab:sigw1} - \ref{tab:sigw3}, the 
cross section
$\sigma_{\gamma p \rightarrow J/\psi Y}$ 
is presented as a function of $\wgp$ for three ranges of $t$ in
the kinematic region $z>0.95$.
The data in each $t$ range are consistent with a power-law dependence 
of the form $\sigma \propto \wgp^{\delta}$ and the results of 
power-law fits for $\delta$ are given in 
table~\ref{tab:slopes}.  
The contribution from correlated systematic errors
is calculated by shifting the
data points according to each source of uncertainty
and repeating the fits.
The values of the power $\delta$ in each $t$ range
are similar to the results from the proton elastic process 
for \jpsi\ mesons at low $|t|$ measured over a similar range of 
$\wgp$~\cite{Adloff:2000vm}.
In a Regge pole model, the power-law dependence
can be expressed as 
${\rm d} \sigma /{\rm d}t = F(t) \wgp^{4\alpha(t)-4}$ 
where $F(t)$ is a 
function of $t$ only. The value of $\alpha(t)$ at each $t$ value
is obtained from $\alpha = (\delta + 4)/4$ 
and is also shown in table~\ref{tab:slopes}. 
Assuming a single effective Pomeron trajectory 
of the linear form 
$\alpha(t) = \alpha(0) + \alpha^{\prime}t$,
a fit to the three $\alpha$ values yields a slope of
$\alpha^{\prime}= -0.0135 \pm 0.0074 \ {\rm (stat.)} 
\pm 0.0051 \ {\rm (syst.)}~{\rm GeV^{-2}}$
with an intercept of 
$\alpha(0) =  1.167 \pm 0.048 \ {\rm (stat.)} \pm 0.024 {\rm (syst.)}$.  
The value of the slope parameter $\alpha^{\prime}$ 
is lower than that observed for the elastic photoproduction of 
\jpsi\ mesons at low $|t|$~\cite{Chekanov:2002xi}.
It is also significantly different from observations at 
low $|t|$ 
in hadron-hadron scattering, 
where a value of 
$\alpha^{\prime}=0.26 \pm 0.02~{\rm GeV^{-2}}$ \cite{Abe:1993xx}
was obtained.

In figure~\ref{fig:w}
the data are compared with the BFKL theoretical 
predictions for the LL approximation (solid curve) and the 
LL+NL prediction with fixed $\alpha_s$ (dashed curve).  
The data are also compared with the DGLAP LL predictions
(dashed-dotted curve).
The BFKL LL
contribution gives a reasonable description of the energy
dependence, except for the lowest $|t|$ range where it is 
steeper than the data. 
The BFKL LL+NL prediction with fixed $\alpha_s$
is similar to that of the BFKL LL prediction.
The DGLAP LL model, which is valid in the range $|t| < M_{J/\psi}^2$,
describes the energy
dependence in the lowest $|t|$ range, $2 < |t| < 5\ {\rm GeV^2}$.
In the region $5 < |t| < 10\ {\rm GeV^2}$, 
where $|t|$ approaches $M_{J/\psi}^2$,
the description becomes worse.

\subsection{Spin Density Matrix Elements}
\label{sect:spin}

The polar ($\theta^{*}$) and azimuthal ($\phi^{*}$) decay 
angular distributions are measured in the rest frame of the \jpsi\
with the quantisation axis taken as the direction of the meson 
in the photon-proton centre-of-mass frame (helicity frame).
The normalised two-dimensional angular distribution for the decay of
the \jpsi\ meson to fermions is written in 
terms of spin density matrix elements
$r_{00}^{04}$, $r_{1-1}^{04}$
and ${\rm Re} \{ {r^{04}_{10}} \}$~\cite{Schilling:1973ag} as
\begin{equation}
\hspace{-4.0cm} \frac{1}{\sigma} 
\frac{{\rm d}^2 \sigma}{{\rm d} \cos \theta^{*} {\rm d}\phi^{*} } = 
\frac{3}{4\pi} \left( \frac{1}{2} (1 + r_{00}^{04}) 
- \frac{1}{2} ( 3r_{00}^{04 } -  1) \cos^2 \theta^{*} \right.  
\label{eq:dthetadphi}
\end{equation}
\[ \hspace{6.5cm} \left. + \sqrt{2}{\rm Re} \{ {r^{04}_{10}} \} \sin 2 \theta^{*} \cos \phi^{*}
+ r^{04}_{1-1} \sin^2 \theta^{*} \cos 2 \phi^{*} \right). \]

The one-dimensional distributions are obtained
by integrating over $\cos \theta^{*}$ or $\phi^{*}$ and give
$\frac{{\rm d} \sigma}{{\rm d} \cos \theta^{*}} \propto 1 + r_{00}^{04} +
(1-3r_{00}^{04})\cos^{2} \theta^{*} $ and
$\frac{{\rm d} \sigma}{{\rm d} \phi^{*}} \propto 1 + r_{1-1}^{04} \cos 2 \phi^{*}$.
Under the assumption of $s$-channel helicity 
conservation (SCHC), the $J/\psi$ meson in photoproduction is expected to be fully
transversely polarised and the matrix elements 
$r_{00}^{04}$, $r_{1-1}^{04}$ and ${\rm Re} \{ {r_{10}^{04}} \}$ are zero.

The spin density matrix elements are extracted by a two-dimensional
log likelihood fit of the data to equation (\ref{eq:dthetadphi}). 
The normalised single differential distributions in $\cos \theta^{*}$
and $\phi^{*}$ are shown in figure~\ref{fig:angular}
for three ranges of $t$.
The dashed curve on the figure shows the expectation from SCHC
and the solid curves show the results of the two-dimensional fit.
The values of the three extracted matrix elements 
are shown in figure~\ref{fig:rrr} and table~\ref{tab:spin} as a function of $|t|$.
Measurements from the ZEUS collaboration  
of the spin density matrix elements for the 
photoproduction of $\rho^0$ and \jpsi\ mesons~\cite{Chekanov:2002rm} 
are also shown in the figure.
In contrast to the $\rho^0$ meson,
the measured spin density matrix elements of the \jpsi\ meson are all
compatible with zero, within experimental errors, and are thus compatible 
with SCHC. 
The \jpsi\ results are therefore
consistent with the longitudinal momentum of the photon being shared 
symmetrically
between the heavy quarks.
Hence, the 
approximations made in the pQCD 
models~\cite{Forshaw:1995ax,Bartels:1996fs,Forshaw:2001pf,Gotsman:2001ne,Enberg:2002zy} for the \jpsi\ wavefunction
are satisfactory for the present data.

\section{Summary}

The differential cross section ${\rm d}\sigma/{\rm d}t$ for the diffractive 
photoproduction of \jpsi\ mesons has been measured 
as a function of the momentum transfer squared $t$ 
from $|t| = 2~\rm GeV^2$ up to values as large as 
$|t|=30~\rm GeV^{2}$ in the kinematic region $z>0.95$ and
$ 50 < \wgp < 150 ~{\rm GeV}$. 
The data are well described in this region by
pQCD calculations\cite{Enberg:2002zy} 
using the leading logarithmic BFKL equation with
parameters consistent with a fit
to  vector meson proton dissociation data at HERA~\cite{Chekanov:2002rm}. 
The addition of non-leading corrections 
preserves the description of the data if the strong coupling $\alpha_s$
is held fixed. The data in the region $|t| < M_{J/\psi}^2$ 
are well described by a model~\cite{Gotsman:2001ne} based on DGLAP evolution.

The cross section has also
been measured as a function of $\wgp$
in three $t$ intervals. The energy dependence shows a similar steep rise 
to that 
observed for elastic \jpsi\ production at low $|t|$~\cite{Adloff:2000vm} 
and the rise persists to the largest $|t|$ values studied. 
The energy dependence is reasonably described by the BFKL model
with the chosen parameters, except for the lowest $|t|$ 
range ($|t| < 5 \ {\rm GeV}^2$).  
The DGLAP model describes the energy dependence in 
the range $|t| < 5 \ {\rm GeV}^2$.

The measurement of the effective Pomeron
trajectory at large $|t|$ yields 
a slope of 
$\alpha^{\prime} = -0.0135 \pm 0.0074 (stat.) \pm 0.0051 (syst.)~{\rm GeV^{-2}}$.
This is lower than that observed for elastic \jpsi\ photoproduction 
at low $|t|$~\cite{Chekanov:2002xi}
and also lower than the slope obtained from hadronic scattering
($\alpha^{\prime} = 0.26 \pm 0.02~{\rm GeV^{-2}}$\cite{Abe:1993xx}).
The observation of the effective slope being small is
compatible with the predictions of models based on 
BFKL evolution~\cite{Bartels:1996fs}.

The spin density matrix elements of the \jpsi\ have been extracted 
in three regions of $t$. 
The results are found to be consistent with $s$-channel 
helicity conservation within the experimental uncertainties
and, therefore, are compatible with models~\cite{Forshaw:1995ax,Bartels:1996fs,Forshaw:2001pf,Gotsman:2001ne,Enberg:2002zy} in which the 
longitudinal momentum of the photon is shared symmetrically
between the quarks of the \jpsi\ .

\section*{Acknowledgements}

We are grateful to the HERA machine group whose outstanding
efforts have made and continue to make this experiment possible. 
We thank the engineers and technicians for their work in constructing and 
maintaining the H1 detector, our funding agencies for 
financial support, the DESY technical staff for continual assistance 
and the DESY directorate for support and for 
the hospitality which they extend to the non-DESY 
members of the collaboration.
We are grateful to R.~Enberg, J.~R.~Forshaw, L.~Motyka, G.~Poludniowski,
E.~Gotsman, E.~Levin, U.~Maor and E.~Naftali for providing
us with the results of their models and for productive discussions.

\newpage

\begin{table}
\begin{center}
\begin{tabular}{|c|c|c|}
\hline
$|t|$ range      & $ \langle |t| \rangle $  & ${\rm d}\sigma/{\rm d} t$ \\   
(${\rm GeV^2}$) & (${\rm GeV^2}$)         & (${\rm nb / GeV^2}$) \\   \hline \hline 
$ 2 -  3 $  & $    2.43 $  & $    5.10  \pm    0.29  \pm    0.65    $ \\ \hline
$ 3 -  4 $  & $    3.45 $  & $    3.08  \pm    0.23  \pm    0.39    $ \\ \hline
$ 4 -  5 $  & $    4.46 $  & $    1.47  \pm    0.15  \pm    0.18    $ \\ \hline
$ 5 -  6 $  & $    5.47 $  & $    0.87  \pm    0.12  \pm    0.11    $ \\ \hline
$ 6 -  7 $  & $    6.47 $  & $   0.610  \pm   0.099  \pm   0.074    $ \\ \hline
$ 7 -  9 $  & $    7.92 $  & $   0.285  \pm   0.046  \pm   0.034    $ \\ \hline
$ 9 - 12 $  & $    10.4 $  & $   0.151  \pm   0.026  \pm   0.017    $ \\ \hline
$12 - 15 $  & $    13.4 $  & $   0.093  \pm   0.020  \pm   0.010    $ \\ \hline
$15 - 21 $  & $    17.7 $  & $  0.0236  \pm  0.0067  \pm  0.0027    $ \\ \hline
$21 - 30 $  & $    25.0 $  & $  0.0045  \pm  0.0023  \pm  0.0005    $ \\ \hline  
\end{tabular}
\caption{The differential photoproduction cross section ${\rm d}\sigma  /{\rm d}t$ 
in the kinematic range $50~<~\wgp<150~{\rm GeV}$ 
and $z>0.95$. The first uncertainty is statistical and the second is systematic.}
\label{tab:dsigdt}
\end{center}
\end{table}

\begin{table}
\begin{center}
\begin{tabular}{|c|c|c|}
\hline
$\wgp$ range      & $ \langle \wgp  \rangle $  & $\sigma_{\gamma p} $ \\   
(${\rm GeV}$) & (${\rm GeV}$)         & (${\rm nb }$) \\   \hline \hline 
$ 50- 68 $  & $    58.4 $  & $    7.26  \pm    0.57  \pm    0.85    $ \\ \hline
$ 68- 86 $  & $    76.5 $  & $    8.11  \pm    0.68  \pm    0.90    $ \\ \hline
$ 86-104 $  & $    94.6 $  & $    9.22  \pm    0.87  \pm    1.06    $ \\ \hline
$104-122 $  & $     113 $  & $    13.5  \pm     1.4  \pm     1.7    $ \\ \hline
$122-140 $  & $     131 $  & $    13.0  \pm     1.8  \pm     1.9    $ \\ \hline
$140-160 $  & $     150 $  & $    14.0  \pm     2.2  \pm     2.4    $ \\ \hline 
 \end{tabular}
\caption{The photoproduction cross section as a function of $\wgp$
integrated over the kinematic range $ 2<|t|< 5 ~{\rm GeV^2}$ 
and $z>0.95$. The first uncertainty is statistical and the second is systematic.}
\label{tab:sigw1}
\end{center}
\end{table}

\begin{table}
\begin{center}
\begin{tabular}{|c|c|c|}
\hline
$\wgp$ range      & $ \langle \wgp \rangle $  & $\sigma_{\gamma p} $ \\   
(${\rm GeV}$) & (${\rm GeV}$)         & (${\rm nb }$) \\   \hline \hline 
$ 50- 82.5 $  & $    64.4 $  & $    1.24  \pm    0.18  \pm    0.14    $ \\ \hline
$ 82.5-115 $  & $    97.4 $  & $    2.75  \pm    0.35  \pm    0.31    $ \\ \hline
$115-147.5 $  & $     130 $  & $    3.98  \pm    0.69  \pm    0.57    $ \\ \hline
$147.5-180 $  & $     163 $  & $    3.26  \pm    0.98  \pm    0.58    $ \\ \hline 
\end{tabular}
\caption{The photoproduction cross section as a function of $\wgp$
integrated over the kinematic range $ 5 < |t| < 10 ~{\rm GeV^2}$ 
and $z>0.95$. The first uncertainty is statistical and the second is systematic.}
\label{tab:sigw2}
\end{center}
\end{table}

\begin{table}
\begin{center}
\begin{tabular}{|c|c|c|}
\hline
$\wgp $ range      & $ \langle \wgp \rangle $  & $\sigma_{\gamma p} $ \\   
(${\rm GeV}$) & (${\rm GeV}$)         & (${\rm nb }$) \\   \hline \hline 
$ 50-100 $  & $    71.0 $  & $   0.499  \pm   0.093  \pm   0.060    $ \\ \hline
$100-150 $  & $     122 $  & $    0.94  \pm    0.19  \pm    0.13    $ \\ \hline
$150-200 $  & $     173 $  & $    1.62  \pm    0.52  \pm    0.38    $ \\ \hline   
\end{tabular}
\caption{The photoproduction cross section as a function of $\wgp$
integrated over the kinematic range $ 10 < |t| < 30 ~{\rm GeV^2}$ 
and $z>0.95$. The first uncertainty is statistical and the second is systematic.}
\label{tab:sigw3}
\end{center}
\end{table}

\begin{table}
\begin{center}
\begin{tabular}{|c|c|c|c|}
\hline
$|t|$ range     & $ \langle |t| \rangle $  &     &  \\   
($\rm GeV^2$) &  (${\rm \ GeV^2}$)  &   
\raisebox{1.5ex}[-1.5ex]{$\delta$}  &  \raisebox{1.5ex}[-1.5ex]{$\alpha$} \\   
\hline \hline
$2-5$     &      
$ 3.06$   &
$0.77   \pm 0.14  \pm 0.10 $ &  
$1.193   \pm 0.035  \pm 0.025   $  \\ \hline
$5-10$  &      
$ 6.93   $   &  
$1.29   \pm 0.23  \pm 0.16  $ &
$1.322   \pm 0.057  \pm 0.040  $ \\ \hline
$10-30$ &      
$ 16.5  $   &
$1.28   \pm 0.39  \pm 0.36  $ &  
$1.322   \pm 0.097  \pm 0.090  $  \\ \hline
\end{tabular}
\caption{The value of $\delta$ obtained when applying a fit to the data of the form
$\sigma (\wgp) \propto {\wgp}^{\delta}$ for each $|t|$ range,
together with the corresponding value of $\alpha$ obtained 
from $\alpha = (\delta + 4)/4$.
The first uncertainty is statistical and the second is systematic.}
\label{tab:slopes}
\end{center}
\end{table}

\begin{table}
\begin{center}
\begin{tabular}{|c|c|c|c|}
\hline
$ \langle |t| \rangle  $  & & & \\
$ ({\rm GeV^2}) $ & \raisebox{1.5ex}[-1.5ex]{$r^{04}_{1-1}$} & 
                    \raisebox{1.5ex}[-1.5ex]{$r_{00}^{04}$} & 
                    \raisebox{1.5ex}[-1.5ex]{${\rm Re} \{ {r^{04}_{10}} \}$} \\  
 \hline \hline
 $ 3.06 $ & $  -0.047  \pm   0.067  \pm   0.009 $  & $   0.01  \pm   0.12  \pm   0.04 $  & $   0.022  \pm   0.069  \pm   0.035    $ \\ \hline
 $ 6.93 $ & $  -0.07  \pm   0.14  \pm   0.07 $  & $  -0.03  \pm   0.17  \pm   0.02 $  & $   0.06  \pm   0.12  \pm   0.05    $ \\ \hline
 $16.5 $ & $  -0.19  \pm   0.22  \pm   0.12 $  & $   0.04  \pm   0.28  \pm   0.04 $  & $  -0.08  \pm   0.19  \pm   0.08    $ \\ \hline
\end{tabular}
\caption{The spin density matrix elements for the
kinematic range $ 50 < \wgp < 150 ~{\rm GeV}$ 
and $z>0.95$. The first uncertainty is statistical and the second is systematic.  The data are quoted at the average $|t|$ values over the ranges given 
in table~\ref{tab:slopes}.}

\label{tab:spin}
\end{center}
\end{table}


\newpage
\setlength{\unitlength}{1pt}

\begin{figure}[h]
\centering
\vskip -1.0cm
\begin{picture}(300,550)
\put(-75,0){\includegraphics[width=17.0cm]{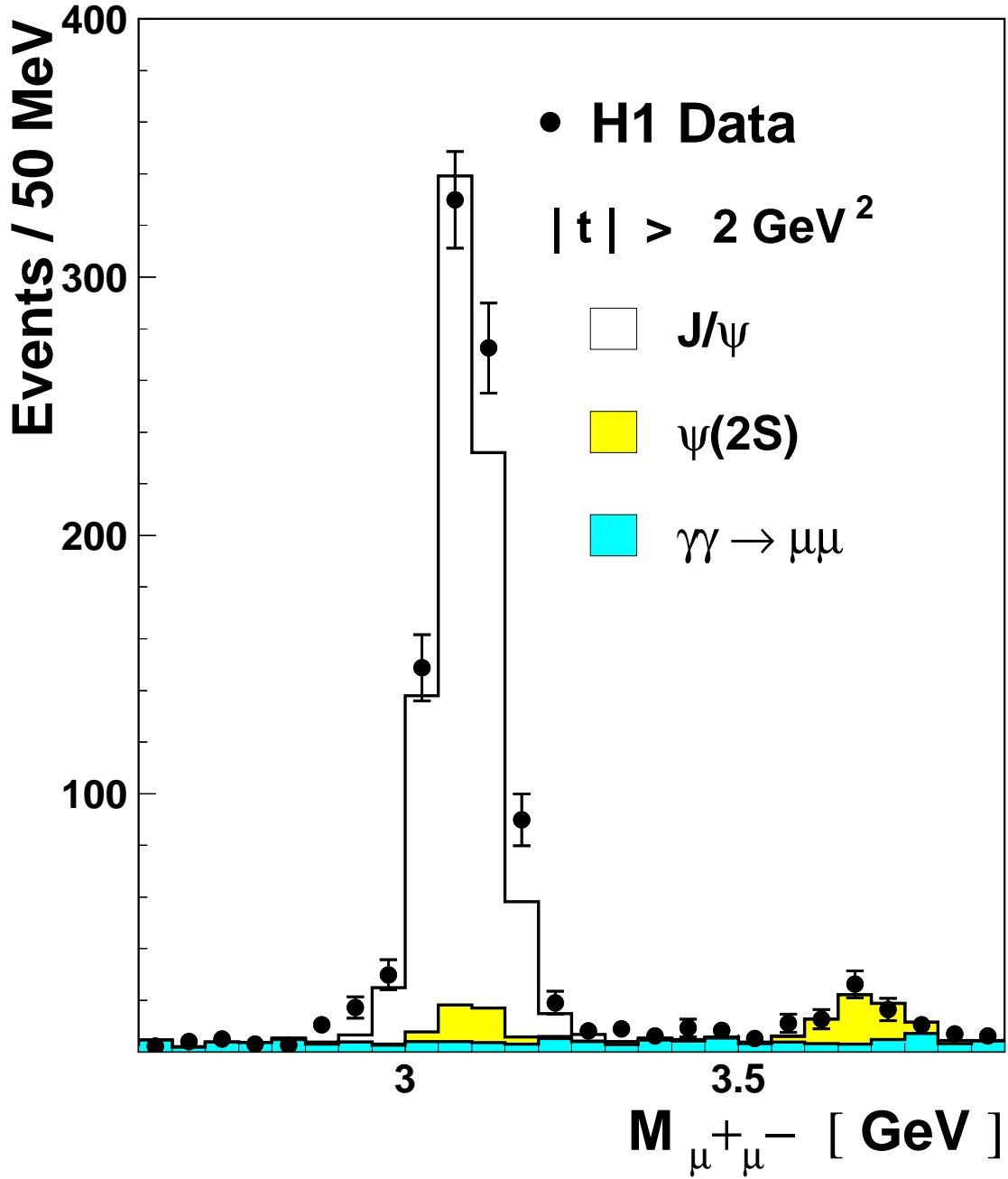}}
\end{picture}
\caption{The $\mu^+ \mu^-$ invariant mass distribution 
in the kinematic region $50 < \wgp <150 \ {\rm GeV}$, $z>0.95$ and
$|t| > 2~{\rm GeV^{2}}$.
The histogram shows the sum of the Monte Carlo simulations 
of $J/\psi$ production using HITVM (open histogram), 
the contribution from lepton pair production as simulated by 
the LPAIR program (dark shaded histogram) and the 
contribution from diffractive $\psi(2S)$ events as simulated with the 
DIFFVM program (light shaded histogram).}
\label{fig:signal}
\end{figure}

\newpage
\begin{figure}[ht]
\centering
\setlength{\unitlength}{1cm}
\begin{picture}(17,19)
\put(-0.8, -.3 ){\includegraphics[height=19.6cm]{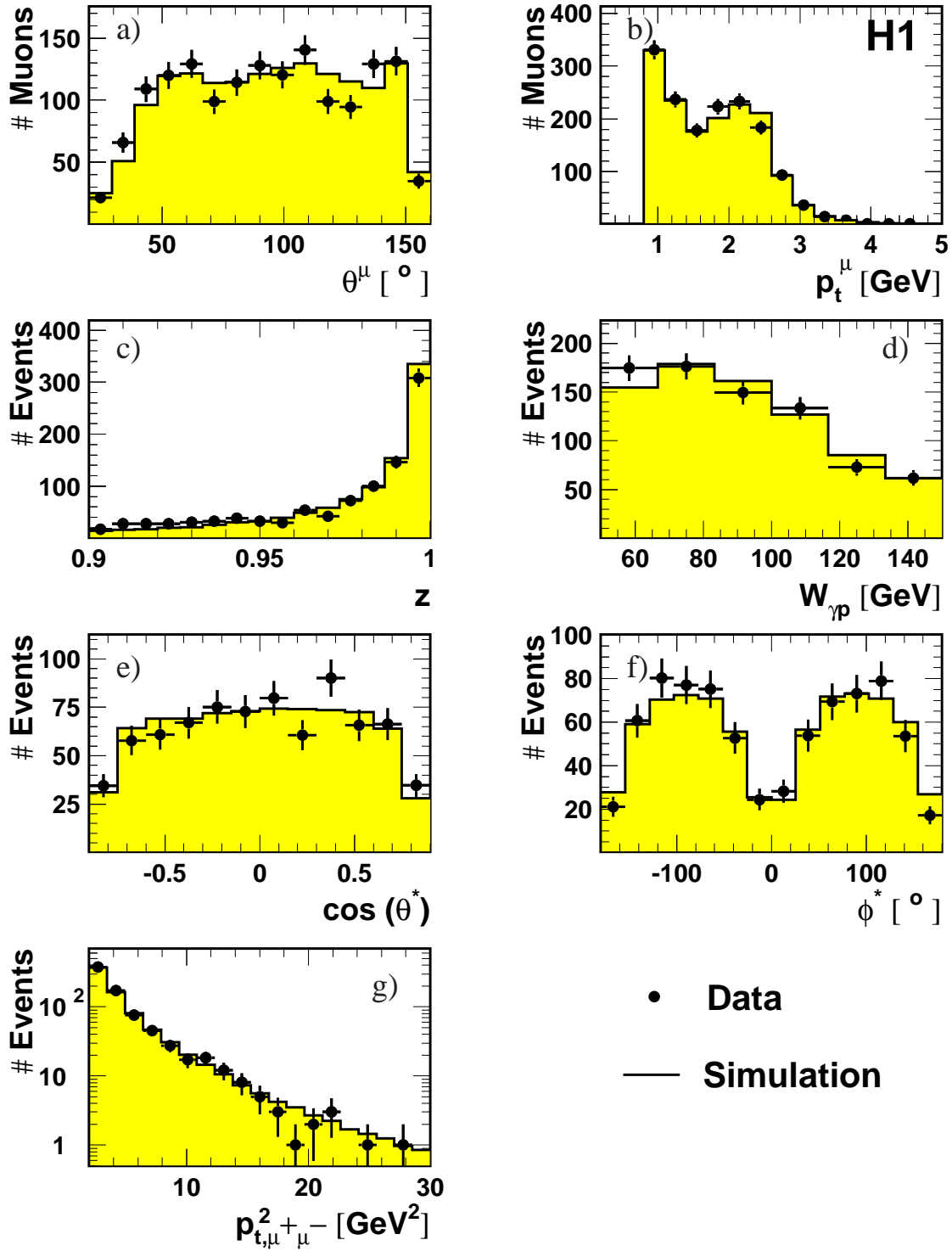}}
\put( 2.3,18.7){\large a)}
\put(10.3,18.7){\large b)}
\put( 2.3,13.7){\large c)}
\put(14.3,13.7){\large d)}
\put( 2.3, 8.7){\large e)}
\put(10.3, 8.7){\large f)}
\put( 6.3, 3.7){\large g)}
\end{picture}
\caption{Kinematic distributions of the dimuon sample in the mass range 
$2.9 < M_{\mu^{+} \mu^{-}} < 3.3 \ {\rm GeV}$.
a) The polar angle $\theta^{\mu}$ and b) the transverse momentum $p_t^{\mu}$
of the muon tracks.
c) The elasticity $z$ and 
d) the photon-proton centre-of-mass energy $\wgp$.
e) The distribution of the cosine of the polar angle 
and f) the azimuthal distribution
of the positively charged decay muon in the helicity frame.
g) The distribution of the squared dimuon transverse momentum.}
\label{fig:control}
\end{figure}

\newpage
\begin{figure}[ht]
\centering
\begin{picture}(400,500)
\put(-25,0){\includegraphics[width=16.5cm]{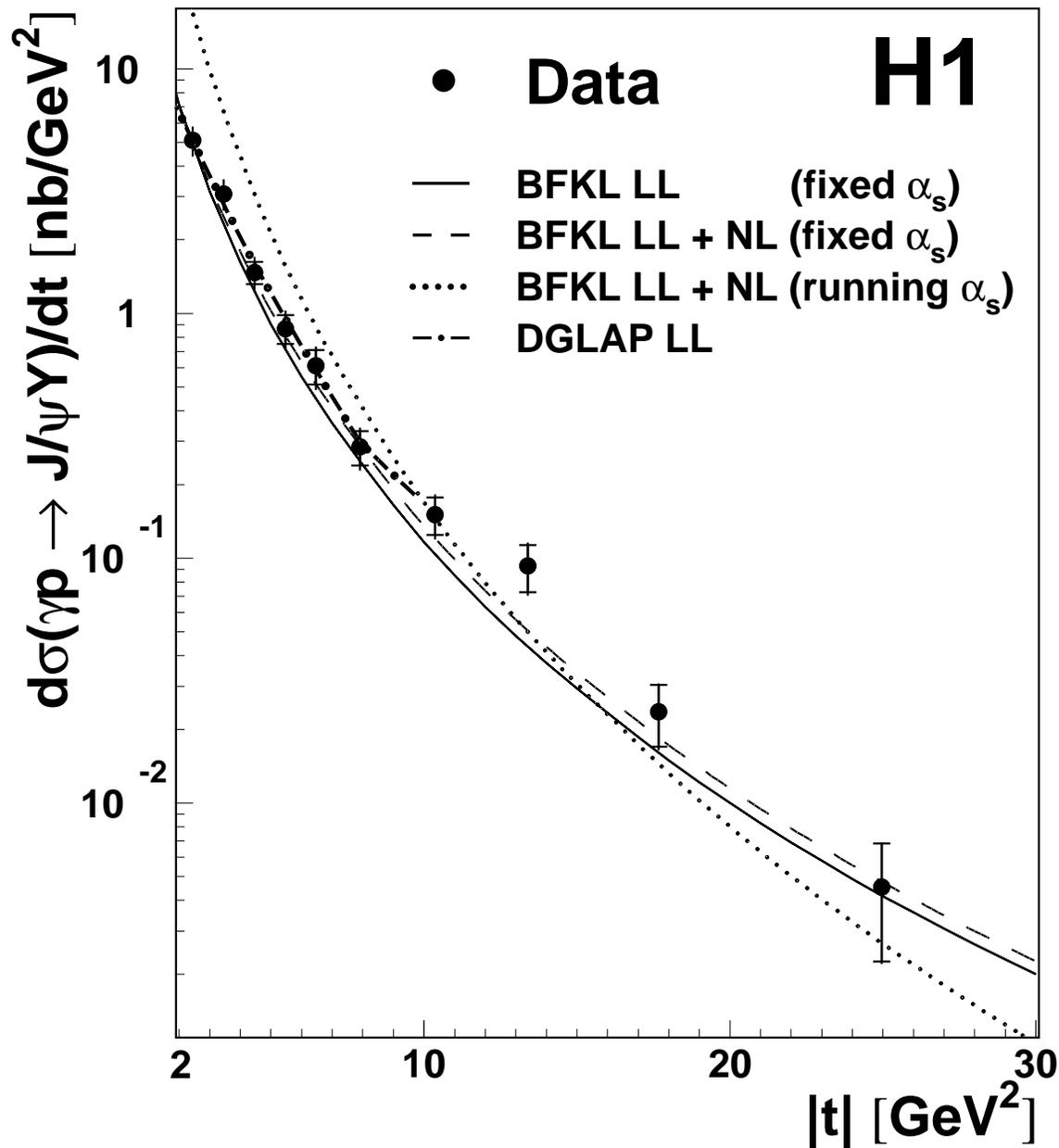}}
\end{picture}
\caption{The photon-proton differential cross section 
${\rm d}\sigma/{\rm d}t$ for $J/\psi$ production
in the kinematic range $50 < \wgp < 150 \ {\rm GeV}$,
$|t|>2.0 \ {\rm GeV}^{2}$ and $z>0.95$. 
The inner error bars correspond to the statistical error and
the outer error bars are the statistical and systematic
errors added in quadrature.
The solid line shows the prediction from the BFKL calculation 
in the leading logarithmic approximation for fixed 
$\alpha_s$~\cite{Enberg:2002zy}.
The dashed (dotted) curve corresponds to the BFKL calculation 
including non-leading corrections and using 
a fixed (running) $\alpha_s$ \cite{Enberg:2002zy}.
The dashed-dotted curve, shown in the range $|t| < M_{J/\psi}^2$,
shows a calculation based on the
DGLAP equation in the leading logarithmic 
approximation \cite{Gotsman:2001ne}.}
\label{fig:t}
\end{figure}

\newpage
\begin{figure}[ht]
\centering
\begin{picture}(400,500)
\put(-25,0){\includegraphics[width=17.0cm]{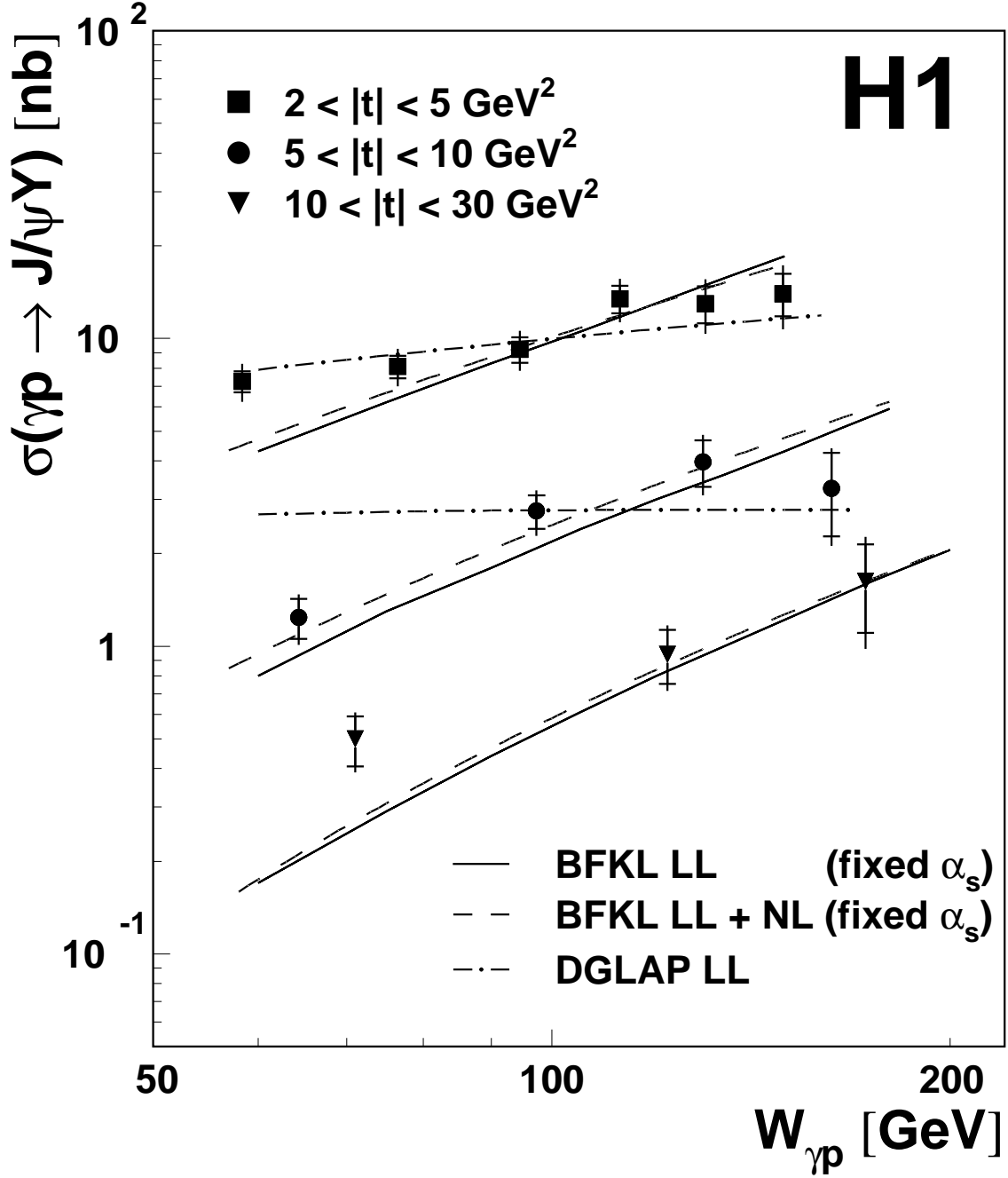}}
\end{picture}
\caption{The photon-proton cross section 
as a function of $\wgp$ in three bins of $|t|$.
The inner error bars correspond to the statistical error and
the outer error bars are the statistical and systematic
errors added in quadrature.
The solid lines show the predictions from the BFKL calculation 
in the leading logarithmic approximation and
the dashed lines correspond to the BFKL calculation 
including non-leading corrections using a 
fixed $\alpha_s$ \cite{Enberg:2002zy}.
The dashed-dotted curve is the result of a calculation based on the
DGLAP equation in the leading logarithmic approximation \cite{Gotsman:2001ne}.}
\label{fig:w}
\end{figure}

\newpage
\begin{figure}[ht]
\setlength{\unitlength}{1cm}
\centering
\begin{picture}(17.,18.)
\put(-0.75,-.5){\includegraphics[width=18cm]{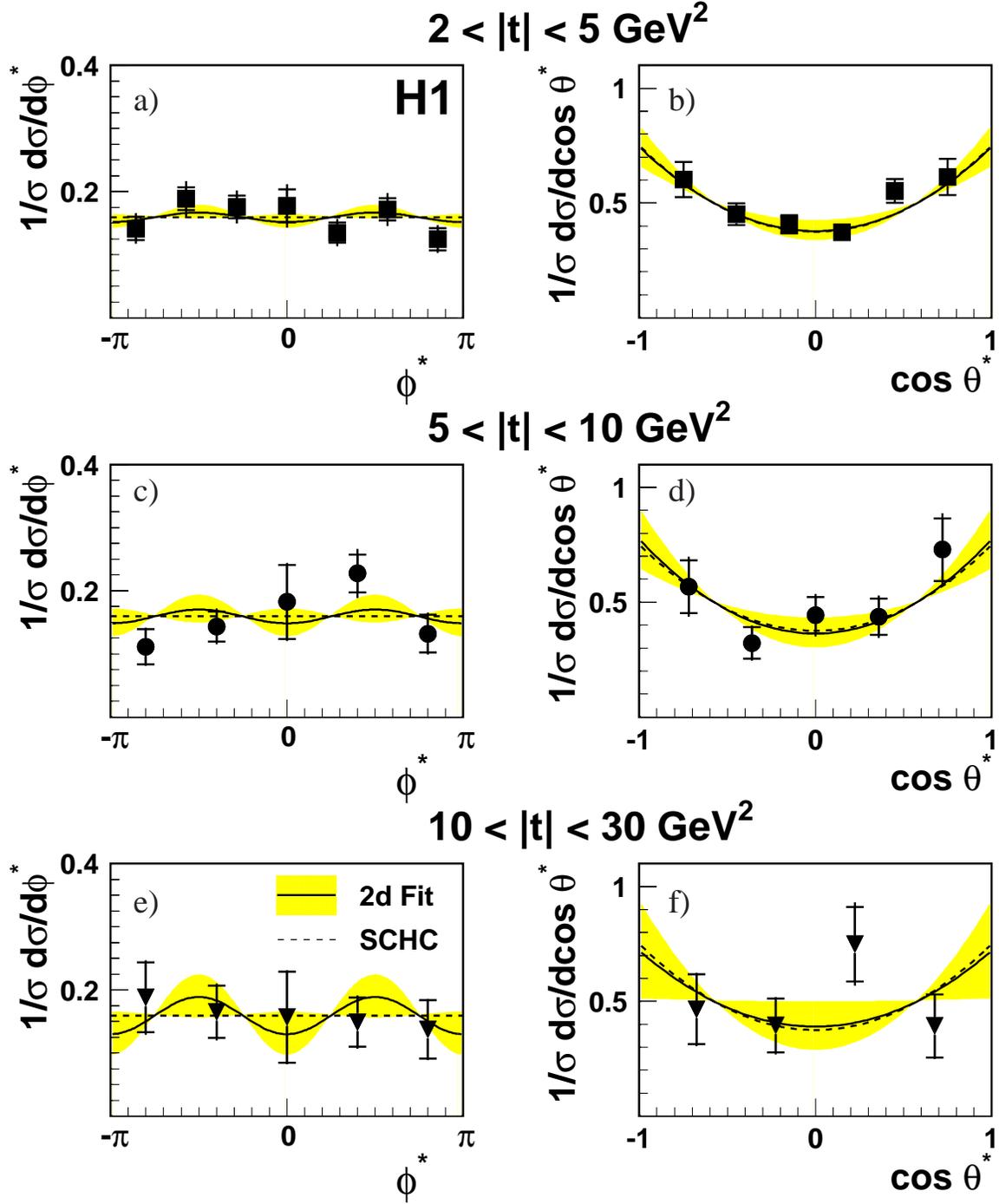}}
\put( 2.3 ,17.0){\large a)}
\put(10.5,17.0){\large b)}
\put( 2.3 ,11.0){\large c)}
\put(10.5,11.0){\large d)}
\put( 2.3 ,4.7){\large e)}
\put(10.5,4.7){\large f)}
\end{picture}
\caption{Normalised decay angular distributions for $J/\psi$ 
meson production in three bins of $|t|$:
a,b) $2<|t|<5 \ {\rm GeV}^2$; c,d) $5<|t|<10 \ {\rm GeV}^2$ and 
e,f) $10<|t|<30 \ {\rm GeV}^2$. 
The left column (a,c,e) shows the azimuthal 
distributions of the positively charged decay muon in the 
helicity frame and the right column (b,d,f) shows the polar 
angle distributions.
The inner error bars show the statistical error and the outer error 
bars show the statistical and systematic errors added in quadrature.
The solid lines show the results of a two-dimensional fit to the 
data (see text).
The shaded band represents the statistical uncertainty for the fit.
The dashed line shows the expectation from $s$-channel helicity
conservation.}
\label{fig:angular}
\end{figure}

\newpage
\begin{figure}[ht]
\setlength{\unitlength}{1cm}
\centering
\begin{picture}(17,18)
\put(-.75,-.5){\includegraphics[width=18.0cm]{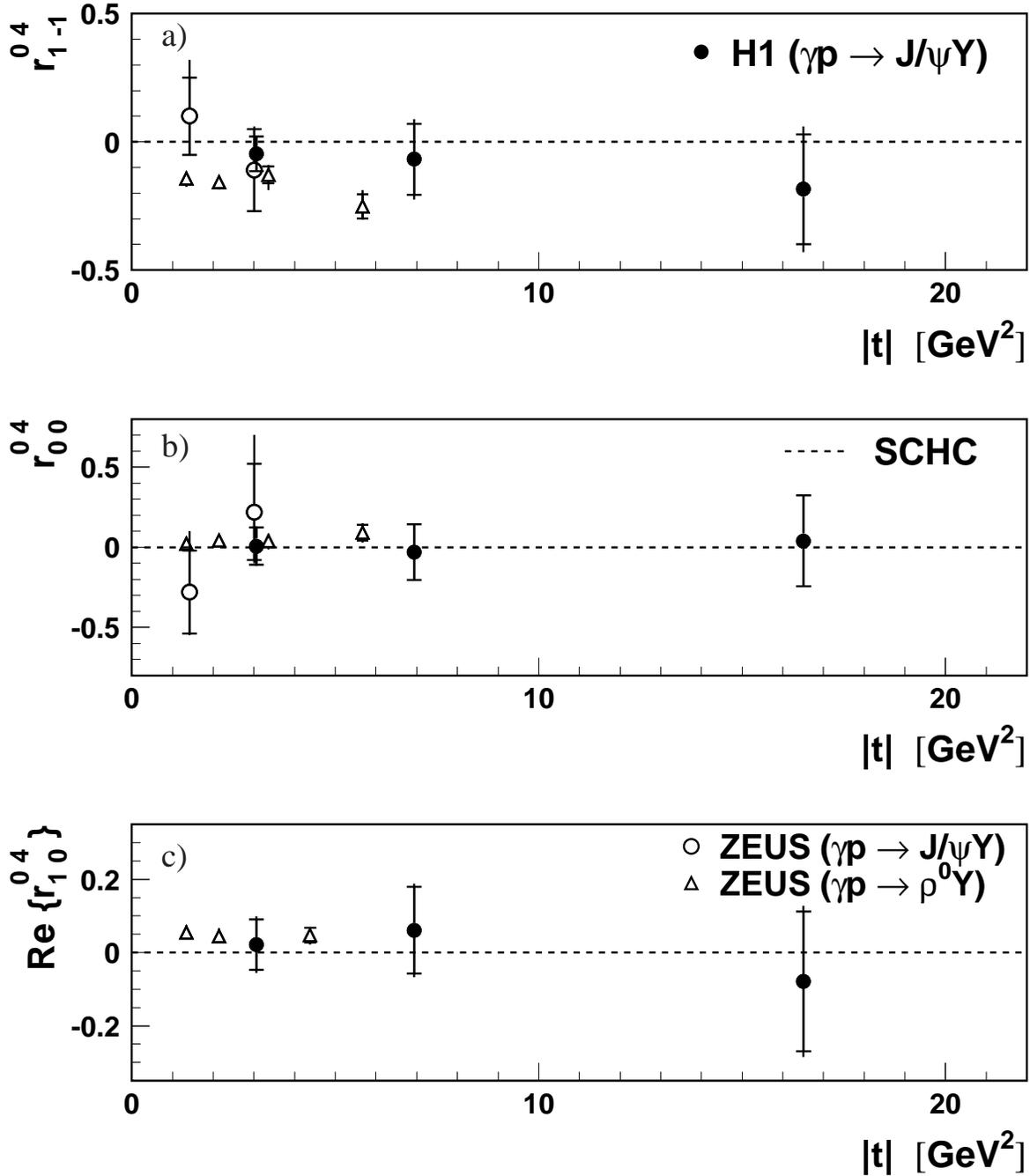}}
\put( 2.4 ,17.2){\large a)}
\put( 2.4, 11.0){\large b)}
\put( 2.4 ,4.8){\large c)}
\end{picture}
\caption{ 
The three spin density matrix elements 
a) $r_{1-1}^{04}$, 
b) $r_{00}^{04}$ 
and c) ${\rm Re} \{ {r^{04}_{10} } \}$
for the \jpsi\ as a function of $|t|$. The inner error bars represent the 
statistical
uncertainty and the outer error bars the combined
statistical and systematic uncertainties.
The dashed line shows the expectation from SCHC. The results from the ZEUS 
collaboration for the
photoproduction of \jpsi\ and $\rho^0$ 
mesons \cite{Chekanov:2002rm} are also shown.}
\label{fig:rrr}
\end{figure}

\end{document}